\documentclass{article}

\usepackage{arxiv}

\usepackage[utf8]{inputenc} 
\usepackage[T1]{fontenc}    
\usepackage{hyperref}       
\usepackage{url}            
\usepackage{booktabs}       
\usepackage{amsfonts,amsmath,amssymb}       
\usepackage{nicefrac}       
\usepackage{microtype}      
\usepackage{lipsum}
\usepackage{graphicx}
\graphicspath{ {./images/} }

\newcommand{\Bad }{objectionability}
\newcommand{\Edu }{educational alignment}
\newcommand{\Read }{readability}

\newcommand{\RankSet}{\textsc{RankSet}}
\newcommand{\ObjSet}{\textsc{ObjSet}}

\newcommand{\ObjCat}{ObjCat}

\newcommand{\BadScore }{$S_{bad}$}
\newcommand{\EduScore }{$S_{edu}$}
\newcommand{\ReadScore }{$S_{read}$}

\newcommand{\MyModel}{\texttt{REdORank}}
\newcommand{\BiGBERT}{\texttt{BiGBERT}}
\newcommand{\ReadFormula}{\texttt{Spache-Allen}}
\newcommand\sa{{\operatorname{\textit{Spache-Allen}}}}

\newcommand{\ObjCLF}{\texttt{$Judge_{bad}$}}

\title{A Multi-Perspective Learning to Rank Approach to Support Children's Information Seeking in the Classroom}

\author{
Garrett Allen \\
  Web Information Systems\\
  Delft University of Technology\\
  Delft, The Netherlands \\
  \texttt{G.M.Allen@tudelft.nl} \\
   \And
 Katherine Landau Wright \\
  Department of Literacy, Language and Culturen\\
  Boise State University\\
  Boise, Idaho, United States \\
  \texttt{katherinewright@boisestate.edu} \\
  \And
 Jerry Alan Fails \\
  Department of Computer Science\\
  Boise State University\\
  Boise, Idaho, United States \\
  \texttt{jerryfails@boisestate.edu} \\
    \And
 Casey Kennington \\
  Department of Computer Science\\
  Boise State University\\
  Boise, Idaho, United States \\
  \texttt{caseykennington@boisestate.edu} \\
   \And
  Maria Soledad Pera\\
  Web Information Systems\\
  Delft University of Technology\\
  Delft, The Netherlands \\
  \texttt{M.S.Pera@tudelft.nl} \\
}

\begin{document}
\maketitle
\begin{abstract}
We introduce a novel re-ranking model that aims to augment the functionality of standard search engines to support \textit{classroom} search activities for \textit{children} (ages 6--11). This model extends the known listwise learning-to-rank framework by balancing risk and reward. Doing so enables the model to prioritize Web resources of high educational alignment, appropriateness, and adequate readability by analyzing the URLs, snippets, and page titles of Web resources retrieved by a given mainstream search engine. Experimental results, including an ablation study and comparisons with existing baselines, showcase the correctness of the proposed model. The outcomes of this work demonstrate the value of considering multiple perspectives inherent to the classroom setting, e.g., educational alignment, readability, and objectionability, when applied to the design of algorithms that can better support children's information discovery.
\end{abstract}


\section{Introduction}
Children in elementary classrooms (Kindergarten--5$^{th}$ grade, typically 6--11 years old) often use search engines (\textbf{SE}) to find Web resources needed to complete their school assignments \cite{azpiazu2017online,rajalakshmi2020bidirectional}. SE built specifically for children's use in a classroom environment, such as EdSearch \cite{LumosLearning} and Kidtopia \cite{Kidtopia}, are known to require regular maintenance. EdSearch manually curates resources (e.g., text or media) to identify those which are educational. Kidtopia instead offers resources from a selection of allow-listed sites using Google's Custom Search (\textbf{GCS}) platform, which utilizes the \textit{SafeSearch} feature to filter out pornographic resources. The allow-listing via manual curation restricts the sites to be both age-appropriate and educational. However, maintaining an up-to-date allow-list becomes burdensome as the Web grows rapidly. Moreover, children's SE based on GCS are known to return less relevant results nearly 30\% of the time, trading relevance for safer results \cite{figueiredo2019false}. Besides these inefficacies, specialized SE must overcome the barrier of adoption: children prefer to use the popular mainstream options for SE, which are known to dominate the market, including Google or Bing~\cite{danovitch2019growing,bilal2019readability,gwizdka2017analysis,azpiazu2017online}.

Mainstream SE are designed and optimized for adults and can overlook unique factors that impact children's use ~\cite{bilal2017towards,madrazo2018looking,vanderschantz2017kids,landoni2021right}. 
This causes barriers to children identifying relevant resources among those presented on a search engine result page (\textbf{SERP}) generated in response to their inquiries \cite{aliannejadi2021children,milton2020korsce}. Children struggle to recognize what and how much information is available online, seldom looking past the first six SERP resources \cite{foss2012children}. Children have trouble understanding the content of retrieved resources due to the complexity of their texts, which leads to uncertainty with relevant resource selection \cite{amendum2016push}. When turning to mainstream SE, children may inadvertently be exposed to inappropriate resources, even when using mainstream functionalities (like Google's \textit{SafeSearch}) as these primarily filter for pornography \cite{zeniarja2018search} and do not account for other potentially harmful content, e.g., violence. Safe search functionality also suffers from over-filtering by preventing resources from being returned if they contain terms that might be mistaken as inappropriate \cite{anuyah2019empirical}.

The research community has allocated efforts to address children's struggles and the shortcomings of current mainstream SE, including methods for sorting search results based on the difficulty of the text, prioritizing websites designed for children, or supporting SERP navigation  \cite{collins2011personalizing,miltsakaki2009matching,gyllstrom2010wisdom,landoni2021right}. Yet, these works share the same quality: they respond to only one of the children's struggles with mainstream SE.

We aim to advance knowledge in the area of Children's Information Retrieval and, more specifically, better enable children's access to online information via SE. As a starting point in our exploration, we focus on tailoring SERP for specific audiences and contexts. To define the scope of our work, we turn to the framework introduced in Landoni et al.~\cite{landoni2019sonny} that allows for the comprehensive design and assessment of search systems for children through four pillars. In our case, these pillars are children aged 6--11 in grades Kindergarten--5$^{th}$ (\textbf{K--5}) as the \textit{user group}, classrooms as the \textit{environment}, information discovery as the \textit{task}, and re-ranking of resources to fit audience and context as the \textit{strategy}.

Within this scope, we pose the following research question (\textbf{RQ}): \textit{Does adapting a learning-to-rank model to account for multiple traits lead to prioritising resources relevant to children and the classroom setting?} We posit that a learning to rank (\textbf{LTR}) strategy can be augmented to simultaneously consider multiple traits of online resources to yield a SERP that prioritizes \textit{educationally valuable} and \textit{comprehensible} resources while minimizing those that are \textit{objectionable}. As such, we introduce \MyModel, a novel re-ranking framework based on multi-perspective LTR meant to support children's use of their preferred SE to complete classroom-related tasks\footnote{This work is based on the MS Thesis of the lead author, which includes an extended discussion of \MyModel, in addition to modules to estimate \Read{}, \Edu{}, and \Bad{} of web resources \cite{allen2021training}.}. This framework leverages the optimization process of LTR to learn a \textit{balance} between the \textit{risks} of inappropriate resources and the \textit{rewards} of contextually relevant resources. In the interest of reproducibility, we share the implementation of \MyModel{} in \texttt{\url{https://github.com/Neelik/REdORank}}.

\MyModel{} is a tangible step towards designing an adaptive search tool for children. We see great potential for using \MyModel{} to support the \textit{searching to learn} portion of the search-as-learning paradigm. Searching to learn is the act of seeking information to gain new knowledge within an educational setting \cite{azpiazu2017online,rieh2016towards}, which aligns very well with the purpose of \MyModel{} given the effectiveness for identifying and propensity to rank higher, Web resources of educational value.

\section{Related Work}
\label{sec:RelatedWork}

When seeking information using mainstream SE, children tend to (i) explore SERP produced in response to their queries using a sequential process from top to bottom and (ii) click higher-ranked results \cite{duarte2011and,foss2012children,gossen2013specifics,gwizdka2017analysis}. As such, it is imperative for SE to prioritize resources that better meet children's needs.

Existing attempts to address this requirement include sorting resources with respect to a user-defined reading level (for middle and high school students) \cite{miltsakaki2009matching}, with the resource's readability calculated using the Coleman-Liau Index \cite{coleman1975computer} together with the LIX and RIX formulas \cite{anderson1983lix}, or re-ranking results matching user reading levels inferred from their search history \cite{collins2011personalizing}. 
Instead, \textit{AgeRank} \cite{gyllstrom2010wisdom}, a modified version of \textit{PageRank}, leverages websites for younger audiences, following the premise that sites designed for children are more likely to link to other child-friendly sites. Iwata et al.~\cite{iwata2010children} consider child-friendliness as part of the re-ranking task, prioritizing resources that are ``easy to understand and visually appealing" for children in elementary school. Syed and Collins-Thompson~\cite{syed2017optimizing} re-rank results for learning utility through an analysis of keyword density, assuming that a user exposed to more keywords in fewer resources will learn information about a subject more successfully. These strategies prioritize resources using a single perspective, which might not be sufficient when serving particular user groups and contexts.

Research on ranking according to education is rich, resulting in strategies based on topic modelling, term clustering, quality indicators, social network attention, and collaborative filtering \cite{premlatha2012re,sanz2010ranking,segal2019difficulty,tanaka2015web,pimentel2018searching}. Notable examples include the work by Marani~\cite{marani2016webedurank}, i.e., \textit{WebEduRank}, who defines a teaching context (a representation of the requirements and experiences of an instructor), to rank learning objects to support instructors. Estivill-Castro and Marani~\cite{estivill2019towards} introduce the Educational Ranking Principle, an algorithm that ranks resources for instructors by analyzing the suitability of a resource for teaching a concept. Acu{\~n}a-Soto et al.~\cite{acuna2019vikor} consider students as part of their audience in their work to rank math videos using a multi-criteria decision-making framework. Unfortunately, as with readability and child-friendliness, children are not the intended user group for most of these works.

Focusing on children in an educational context, Yilmaz et al.~\cite{yilmaz2019improving} introduce a strategy to automatically label queries that align with educational subjects. These predicted labels are incorporated as an indicator for re-ranking resources. Like \MyModel, this ranker incorporates an education alignment but is centred on the Turkish education system and the Turkish language. Usta et al.~\cite{usta2021learning} train an LTR model for a query-dependent ranking strategy aimed at prioritizing educational resources for students in the 4$^{th}$--8$^{th}$ grades. Through feature engineering, the authors extract disjointed sets of features from the query logs of a Turkish educational platform called Vitamin \cite{usta2014k}: (i) query-document text similarity, (ii) query specific, (iii) document specific, (iv) session based, and (v) query document click based. Unique to this approach is that within the query- and document-specific groups are domain-specific features such as course, grade, and document type, e.g., lecture, video, or text. This approach differs from ours in that the features used in training the ranker originate from a domain-specific SE that includes course and grade information of the resources. In contrast, we design a re-ranker that is SE agnostic, allowing our re-ranker to be coupled with any generic SE. Additionally, the features used by Usta et al.~\cite{usta2021learning} include click data originating from children, which is not readily or publicly available for our user group.

\begin{figure*}[ht]
    \centering
    \includegraphics[width=0.95\textwidth]{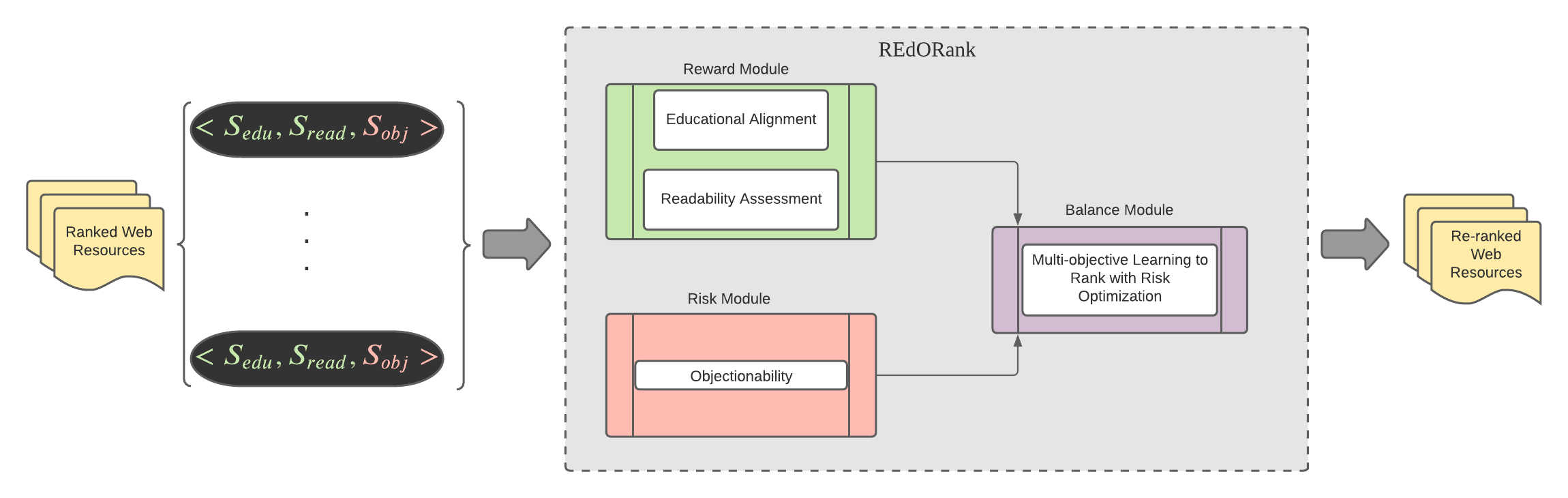}
    \caption{The \MyModel{} framework, which re-ranks SE resources retrieved in response to a child's classroom-related query balancing reward with risk.}
    \label{fig:my-framework}
\end{figure*}

A strategy closely related to \MyModel{} is \textit{Korsce} \cite{milton2020korsce}, which also examines the appropriateness, curriculum alignment, objectivity, and reading comprehensibility of resources to identify those that best match 3$^{rd}$ -- 5$^{th}$ grade children searching in the classroom. Korsce treats resources as inappropriate if they refer to pornography and hate speech but fails to account for other potentially objectionable topics like alcohol or drugs. For curriculum alignment, Korsce uses topic modelling based on Latent Dirichlet Allocation, which follows a word-level and semantic space exploration of resources but does not account for the contextual information that can be garnered from considering resource text in its entirety. 

For reading comprehension estimation, Korsce relies on the Flesch-Kincaid formula and a cosine curve that penalizes resources with a readability level exceeding the expected grade level. However, other formulas have been reported to be more effective when predicting the readability levels of K--12 resources \cite{allen2022super}. Further, Korsce requires the expected grade for the user, which is rarely available for mainstream SE.

Korsce ranks resources according to a static set of optimal weights, manually chosen as the result of empirical exploration of near-optimal rankers \cite{van2016balancing}. The selected ranker generates scores resource-by-resource (akin to pointwise methods). Alternatively, we utilize a listwise approach, allowing for absolute relevance comparisons among resources, as all resources are considered simultaneously instead of independently.

\section{Methodology}
\label{ch:method}

\MyModel{} is a multi-perspective learning to rank framework that re-ranks resources through examining \textit{in tandem} the \underline{R}eadability, \underline{Ed}ucational alignment, and \underline{O}bjectionability of each resource $R$ retrieved by a mainstream SE in response to a child's query inquiring on classroom-related concepts. Taking advantage of the retrieval power of mainstream SE, \MyModel{} identifies and prioritizes resources intended for K--5 classrooms and students.
As shown in Figure \ref{fig:my-framework}, \MyModel{} consists of three modules: the \textit{reward} module, the \textit{risk} module, and a \textit{balance} module.

\subsection{Reward}
The reward module determines the interaction between ``positive" perspectives for resource analysis: \Read{} and alignment with the classroom curriculum. 

\textbf{Readability.} For resources to be useful, children must be able to decode and comprehend the information within them. Children who read above their reading level experience lower reading comprehension \cite{amendum2018does}. Readability, or ``the overall effect of language usage and composition on a readers' ability to easily and quickly comprehend the document" \cite{meng2020readnet}, aids in identifying resources that children can understand. However, estimating grade levels of online resources is not simple, given the broad range of formulas available for estimation purposes. In addition, no consensus exists on which formula should be used for online resources. Allen et al.~\cite{allen2022super} introduced a formula, \ReadFormula{}, that utilizes a large vocabulary comprised of a broad range of terms that children acquire as they age to determine the readability of a text. \ReadFormula{} was empirically found to be effective for estimating the reading difficulty of children's online resources, which we leverage in \MyModel.

As shown in Eq. \ref{eq:readability-score}, the readability score \ReadScore{} of $R$, inferred using its snippet $R_{S}$\footnote{Due to the complexities of gathering, computing resources, and storage needs for processing the full content of Web pages, we use snippets as a proxy for the full page content.} is determined by \ReadFormula.

\begin{equation}
\label{eq:readability-score}
    S_{read}(R) = \sa(R_{S})
\end{equation}

\textbf{Educational Resources.} Not all resources aligned with the reading abilities of children are suitable for the classroom. To explicitly respond to our environment, \MyModel{} considers the \Edu{} of resources and aims to promote those with educational value, as previous research has shown that ranking educational resources higher in search results has the potential to increase learning efficiency \cite{syed2017optimizing}. In our case, educationally aligned resources are defined as those that align with established educational guidelines that provide a set of learning outcomes for each grade that K--12 students are expected to achieve. 

In particular, we focus on educational resources that inform on subjects for grades K--5, such as language arts, science, and social studies. As shown in Eq. \ref{eq:educational-score}, to capture the degree to which a web resource $R$ is educationally aligned, we employ \BiGBERT \cite{allen2021bigbert}, the \underline{Bi}directional \underline{G}ated Recurrent Unit with \underline{BERT} model. \BiGBERT{} examines the URL ($R_{U}$) and snippet ($R_{S}$) of $R$ based on known educational standards, such as the United States' Common Core State Standards and the Next Generation Science Standards. \EduScore{} has a range of $[0,1]$. 

\begin{equation}
\label{eq:educational-score}
    S_{edu}(R) = BiGBERT(R_{S}, R_{U})
\end{equation}

\subsection{Risk}
The risk module looks at the interaction of ``negative" perspectives that identify resources as inappropriate for the user group. 

\textbf{Objectionable Resources.} The Web contains an ever-growing collection of resources for users of many ages, experiences, and knowledge levels. It is therefore anticipated that some of these resources are more attuned to some user groups than others. Given the user group and environment that is the focus of this work, it is critical for \MyModel{} to mitigate the risk of presenting resources towards the top of SERP that could be deemed inappropriate. Preventing the display of inappropriate results while also avoiding over-filtering results that may appear as objectionable but are not, e.g., an article on breast cancer \cite{figueiredo2019false}, requires a solution that goes beyond safe search. To account for the large variety of objectionable material present online, and inspired by prior strategies to detect objectionable resources \cite{milton2020korsce,lee2013objectionable}, we treat as objectionable for children in the classroom resources that relate to any category in \ObjCat: Abortion, Drugs, Hate Speech, Illegal Affairs, Gambling, Pornography, and Violence. Note that the Drugs category refers to resources over-arching \textit{drugs}, but also \textit{alcohol}, \textit{tobacco}, and \textit{marijuana}. Further, Violence focuses on violent content, as well as \textit{weapons}; Hate Speech accounts for \textit{racism} and hateful/offensive content.

To determine whether resources are likely to be objectionable, we build upon the state of the art to produce \ObjCLF, a lexicon-based classification model that scrutinizes their terminology. This requires the existence of pre-defined lists of `objectionable' terms. In the case of the Pornography and Hate Speech categories, we use the pre-defined lists used in \cite{milton2020korsce}, which are sourced from Google's archive \cite{googleBadWords} and the Hate Speech Movement's website \cite{hsm}, resp. Unfortunately, there are no curated term lists associated with the remaining categories in \ObjCat. Thus, we generate them through a novel process called category understanding via label name replacement \cite{meng2020text}.

We use websites from Alexa Top Sites \cite{amz2020alexa} known to belong to categories appearing in \ObjCat{} as our corpus for generating the term lists. For each category, excluding Pornography and Hate Speech, the occurrence of the category name (as well as sub-category names, if available) within a website from the corpus is masked, and a pre-trained BERT encoder is used to produce a contextualized vector representation $h$ with the masked category name. BERT's masked language model head produces a probability distribution that a term $w$ from within BERT's vocabulary will occur at the location of the masked category name. 

Terms can occur in different contexts within the same corpus. Thus, terms in the extracted vocabulary are ranked by their probability of occurrence (Eq. \ref{eq:meng-category-vocab-prediction}) and by how many times each term can replace a category name in the corpus while maintaining context. 

\begin{equation}
\label{eq:meng-category-vocab-prediction}
    p(w\: |\: h) = Softmax\:(W_{2}\: \sigma\: (W_{1}h + b))
\end{equation}
\noindent
where $\sigma(\cdot)$ is the activation function; $W_{1}$, $W_{2}$, and $b$ are learned parameters for the masked language prediction task, pre-trained within BERT.

As in \cite{meng2020text}, we select the top 100 terms per category (or the entire list if less than 100 are extracted) as the representative term list that captures contextually similar and synonymous terms associated with the corresponding categories.

We represent $R$ with a collection of 16 text-based features extracted from its snippet $R_{S}$. Seven of these account for the prevalence (i.e., frequency of occurrence) of objectionable terms in $R_{S}$. A further seven features account for scenarios where a term could be misconstrued as objectionable depending on context. We also consider that producers of objectionable online content are known to introduce intended misspellings as an attempt to bypass safe search filters \cite{milton2020korsce}, and therefore capture the prevalence and coverage of misspellings.

For each category $oc$ in \ObjCat, we calculate the term prevalence, i.e., $TP(R_{S}, oc)$, as in Eq. \ref{eq:term-prevalence}.

\begin{equation}
\label{eq:term-prevalence}
    TP(R_{S}, oc) = \frac{\sum_{t \in TL_{oc}} tf(t, R_{S})}{|R_{S}|}
\end{equation}
\noindent
where $TL_{oc}$ is the term list for $oc$, $t$ is a term in $TL_{oc}$, and $tf(\cdot)$ is a function that calculates the number of times $t$ appears in $R_{S}$. Serving as a normalization factor, $|R_{S}|$ is the length of $R_{S}$ after tokenization, punctuation \& stop word removal, and lemmatization (using the NLTK Python library).

We also consider the coverage of objectionable terminology in $R_{S}$ as, for example, ``breast" could frequently occur in a biology resource that is itself not objectionable; it can also appear in a pornographic resource. For each category $oc$, we calculate objectionable term coverage in $R_{S}$, i.e., $TCov(R_{S}, oc)$, using Eq. \ref{eq:obj-term-coverage}.

\begin{equation}
\label{eq:obj-term-coverage}
    TCov(R_{S}, oc) = \frac{\sum_{t \in TL_{oc}} \delta\:(t, R_{S})}{|\:TL_{oc}\:|}
\end{equation}
\noindent
where $TL_{oc}$ and $t$ are as defined in Eq. \ref{eq:term-prevalence}, $\delta(t, R_{S})$ is 1 if $t$ occurs at least once in $R_{S}$ and 0 otherwise, and $|TL_{oc}|$, the total number of terms in $TL_{oc}$, acts as a normalization factor.

We explicitly account for misspelled terms by looking at their prevalence in $R_{s}$--how often misspellings occur in $R_{s}$--using Eq. \ref{eq:misspelled-term-prevalence}. 

\begin{equation}
\label{eq:misspelled-term-prevalence}
    MP(R_{s}) = \frac{\sum_{t \in R_{S}} \beta\:(t, R_{s})}{|R_{S}|}
\end{equation}
\noindent
where $t$ is a term in $R_{S}$, $\beta(t, R_{S})$ is 1 if $t$ is a misspelling and 0 otherwise, and $|R_{S}|$ is a normalization factor representing the length of $R_{S}$. We use the Enchant library \cite{Enchant} to identify misspelled terms as it wraps many existing spellchecking libraries, such as Ispell, Aspell, and MySpell.

Lastly, we look at the coverage of misspellings using Eq.~\ref{eq:misspelled-term-coverage}.

\begin{equation}
\label{eq:misspelled-term-coverage}
    MC(R_{s}) = \frac{\sum_{t \in R_{Su}} \gamma\:(t, TL_{all})}{\sum_{t \in R_{Su}} \beta\:(t, R_{s})}
\end{equation}
\noindent where $\beta(.)$ is defined as in Eq. \ref{eq:misspelled-term-prevalence}, $t$ is a term in $R_{Su}$, which is the set of unique terms in $R_{S}$, $TL_{all}$ is the set of terms resulting from merging the term list for each category in \ObjCat, and $\gamma(.)$ evaluates to 1 if $t$ is identified as a misspelling and it occurs in $TL_{all}$, and 0 otherwise.

Based on its effectiveness in similar classification tasks \cite{milton2020korsce}, we use the Random Forest model to identify objectionable resources. Using the feature representation of $R$ as input, a trained Random Forest model\footnote{Max leaf node, min leaf samples, and min sample split are set to 32. Max depth is set to 8.} produces a binary probability distribution \textbf{\^y} over each class--objectionable and not--such that \^y $\in{[0,1]}$ for $R$. To serve as the sensitivity score exploited by the risk module, we define \BadScore{} as the probability value of $R$ being associated with the objectionable class (Eq. \ref{eq:objectionable-score}).

\begin{equation}
\label{eq:objectionable-score}
    S_{bad}(R) = Judge_{bad}(R_{S})
\end{equation}

\subsection{Balance}
The balancing module trades off outputs of the risk module (a value that acts as cost and therefore decreases resource prioritization) and the reward module (a value meant to increase resource prioritization in the ranking), resulting in a final ranking score by which resources are reordered.

\textbf{Listwise LTR}. LTR is a machine learning strategy that, when applied to Information Retrieval, refers to the task of automatically constructing ``a ranking model using training data, such that the model can sort new objects according to their degrees of relevance, preference, or importance'' \cite{liu2011learning}. Advancements in LTR models have expanded the loss function to accept more than one resource as input, resulting in the following categorizations for LTR models: pointwise, pairwise, or listwise \cite{li2011short}. These variations are based on whether a single resource, a pair of resources, or a list of resources, respectively, are operated over during the optimization of the loss function.

When used for Web search, models using listwise loss functions have been shown to be more effective in terms of ranking accuracy and degree of certainty of ranking accuracy in relation to the pointwise and pairwise counterparts \cite{cao2007learning,tax2015cross}. Well-known listwise models \cite{cao2007learning,dai2020u,ma2019online,pang2020setrank,xia2008listwise,xu2007adarank}, however, optimize their respective ranking functions on a single relevance measure.

In practice, the relevance of a search result is not always established based on a single trait (particularly as relevance judgments are often associated, among others, with concepts like usefulness, utility, and pertinence \cite{borlund2003concept,schamber1990re}). For instance, students searching for information on John Adams for a class assignment would determine resource relevance by considering factors such as whether a resource uses language they can understand, whether the John Adams being discussed is the correct individual, and whether the resource discusses the aspect of John Adams for which they are seeking information, i.e., information on his term as President vs. information on his role during the American Revolution. To better align with such real-world scenarios, multi-objective LTR strategies that optimize loss functions for multiple measures of relevance have been brought forth \cite{bruch2020stochastic,svore2011learning,van2016balancing}. Yet, such approaches opt for the pairwise variation of LTR \cite{carmel2020multi,momma2020multi}. 

\textbf{AdaRank}. When accounting for multiple objectives, listwise approaches like AdaRank are rarely considered. AdaRank \cite{xu2007adarank} is one of the more prevalent algorithms in LTR research~\cite{geetha2019knowledge,kuzi2019analysis,lie2018novel,mcbride2019cost}. AdaRank uses a listwise approach, which is the most effective in terms of ranking accuracy when used for Web search \cite{cao2007learning,tax2015cross}. AdaRank learns a ranking function through the optimization of an evaluation measure. The metric most commonly used for optimization is Normalized Discounted Cumulative Gain (\textbf{NDCG}) \cite{jarvelin2017ir,liu2011learning}. The goal of NDCG is to measure the agreement between a predicted ranked list and the ground truth for a query. However, this style of LTR is geared towards a single relevance value with respect to a query and does not account for any sort of ``risk" factor of resources.

\textbf{Cost-sensitive Optimization}. The goal of a search system is to retrieve resources from a collection that have the highest relevance with regard to a user's query. In some cases, these collections contain resources that are not meant to be seen by all users, such as private medical documents or, in the case of a government system, top secret missives. These types of resources are known as sensitive resources. To avoid presenting sensitive materials in response to online inquiries, Sayed and Oard~\cite{sayed2019jointly} introduced an extended version of the DCG metric, called Cost Sensitive Discounted Cumulative Gain (\textbf{CS-DCG}). This new metric (Eq. \ref{eq:CS-DCG}) introduces a cost penalty, or a risk factor, for displaying a sensitive document within a ranking of retrieved resources.

\begin{equation}
\label{eq:CS-DCG}
    CS-DCG_{k} = \sum_{i=1}^{k} \frac{g_{i}}{d_{i}} - c_{i}
\end{equation}
\noindent where $k$ is a cutoff value, i.e., the number of resources examined in a list and $i$ is a position in the ranking. $g_i$ is the relevance gain of the $i^{th}$ resource, and $d_i$ is the discount for the $i^{th}$ resource.

Incorporating CS-DCG into an LTR model such as AdaRank empowers the model to learn to rank sensitive documents lower than those that are not sensitive. This aligns with what we seek to do with the \Bad{} perspective of \MyModel: eradicate from top-ranking positions resources that can be perceived as sensitive for the user group and environment that are the focus of our work. Thus, instead of depending upon the traditional NDCG for optimizing its LTR re-ranker, \MyModel{} uses CS-DCG. In this case, we use as the sensitivity cost $c_{i}$, \BadScore{} (Eq. \ref{eq:objectionable-score}). 

CS-DCG accounts for objectionable resources but still only considers a single signal for relevance gain. In the context of our work, however, it is imperative to leverage the influence that both \Edu{} and \Read{} have into determining the relevance of a given resource. It is not sufficient to simply linearly combine the respective grade level and educational alignment scores, \EduScore{} and \ReadScore. Instead, it is important to understand the interdependence between these two scores in terms of dictating relevance gain.

To model the connection between \Edu{} and \Read{}, we take inspiration from a weighting scheme core to Information Retrieval: TF-IDF. TF (term frequency) captures the prominence of a term within a resource, whereas IDF (inverse document frequency) characterizes the ``amount of information carried by a term, as defined in information theory" \cite{croft2010search} and is computed as a proportion of the size of a collection over the number of resources in the collection in which the term appears. In our case, this weighting scheme acts as a sort of ``mixer" for the traits that inform relevance. Intuitively, we treat \EduScore{} as representative of the content of $R$ (in terms of matching the classroom setting) and \Read{} as the discriminant factor with respect to resources considered for ranking purposes. Given the often high readability levels of online resources \cite{antunes2019readability,anuyah2019empirical}, we use 13 as the readability level representative of the collection and therefore use it as the max \Read{} in the numerator for IDF. With this in mind, the mixer score for $R$ informed by the two aforementioned signals of relevance is computed as in Eq. \ref{eq:mixer}.

\begin{equation}
\label{eq:mixer}
    mixer(R) = S_{read}(R) \times log_2 (\frac{13}{S_{edu}(R)})
\end{equation} 

By incorporating multiple signals of relevance into the determination of relevance gain, and the expansion of DCG with a cost-sensitivity factor, we have defined an updated metric that ensures \MyModel{} explicitly learns to respond to the user group, task and environment requirements by prioritizing resources that align with our user group and environment, while preventing the presentation high in the ranking of resources that are objectionable for our environment.

\section{Experimental Set-up}
\label{sec:Eval-Ranking}
In this section, we discuss the extensive experiments outlined to validate the design of \MyModel{}.

\begin{table*}[ht!]
\centering
\caption{Performance of \MyModel{} and ablation variations using \RankSet. The suffixes -R, -E, -O indicate Readability only, Educational only, and Objectionable only, resp. -M indicates the use of the mixer for \Edu{} and \Read; -MER shows the use of the mixer \textit{with} -E and -R. * indicates significance w.r.t. \MyModel{} and bold indicates best performing for each metric.}
\label{tab:ranking-results-ablation}
{%
\begin{tabular}{@{}clllll@{}}
\toprule
\multicolumn{1}{l}{\textbf{Row}} & \multicolumn{1}{l}{{\textbf{Algorithm}}} & \multicolumn{1}{l}{\textbf{\begin{tabular}[l]{@{}l@{}}Optimization \\ Metric\end{tabular}}} & \multicolumn{1}{l}{\textbf{NDCG}} & \multicolumn{1}{l}{\textbf{\begin{tabular}[l]{@{}l@{}}MRR\end{tabular}}} & \multicolumn{1}{l}{\textbf{\begin{tabular}[l]{@{}l@{}}MRR$_{Bad}$\end{tabular}}} \\ \midrule
1                       & AdaRank                                                                               & NDCG                                                                                          & 0.778*                            & 0.226*                                                                               & 0.097*                                                                                  \\
2                           & AdaRank-E                                                                             & NDCG                                                                                          & 0.765*                            & 0.209                                                                               & 0.110*                                                                                  \\
3                           & AdaRank-R                                                                             & NDCG                                                                                          & 0.774*                            & 0.222                                                                               & 0.101*                                                                                  \\
4                           & AdaRank-O                                                                             & NDCG                                                                                          & 0.675*                            & 0.148*                                                                              & 0.537*                                                                                  \\
5                           & REdORank-E                                                                            & nCS-DCG                                                                                       & 0.765*                            & 0.209                                                                               & 0.110*                                                                                  \\
6                           & REdORank-R                                                                            & nCS-DCG                                                                                       & 0.774*                            & 0.222                                                                               & 0.101*                                                                                  \\
7                           & REdORank-O                                                                            & nCS-DCG                                                                                       & 0.675*                            & 0.148*                                                                              & 0.537*                                                                                  \\
8                           & REdORank-M                                                                            & nCS-DCG                                                                                       & 0.765*                            & 0.209                                                                               & 0.110*                                                                                  \\
9                           & REdORank-MER                                                                            & nCS-DCG                                                                                       & 0.777                            & 0.218                                                                               & \textbf{0.089}*                                                                                  \\
10                           & REdORank                                                                              & nCS-DCG                                                                                       & \textbf{0.779}                    & \textbf{0.228}                                                                      & 0.097                                                                                   \\\bottomrule
\end{tabular}%
}
\end{table*}

While \textbf{datasets} like MQ2007 \cite{qin2013introducing} or OHSUMED \cite{hersh1994ohsumed} are available for evaluating models based on LTR, none is comprised of queries, resources, and ``ideal" labels pertaining to our user group and environment. In addition, none of these datasets includes known objectionable resources, which are crucial for explicitly assessing the validity of \MyModel's design. Thus, we construct our own dataset: \RankSet. 
The construction of datasets for ranking tasks in information retrieval often follows the Cranfield paradigm \cite{voorhees2019evolution}. This process involves beginning with known ``ideal" resources. The title of each resource is used as a query to trigger the retrieval of other resources to produce a ranked list. The ideal resource is always positioned at the top of the ranking, as it is treated as the ground truth. The remaining top-N ranked resources (excluding the one originating the search, if available) are used to complete the ranked list. Following this paradigm, we create \RankSet{} and ensure an ideal resource is in the top position for every query. However, \MyModel{} also aims to push objectionable resources lower in the rankings. To enable evaluation of this aspect of \MyModel, we append at the bottom of the list a known ``bad" resource.
 
To act as the ideal resources for \RankSet, we use a collection of 9,540 articles with known reading levels and educational value targeted for children on a variety of topics from NewsELA \cite{newsELA}. For bad resources, we turn to \ObjSet{}. Following the Cranfield paradigm, we use the ideal article titles as queries and using Google's API, we retrieve up to 20 resources, their titles, search snippets, and rank positions (we drop queries that lead to no resources or resources with missing content). We assign relevance labels of 2 to the ideal resources, 0 to the known ``bad" resources, and 1 to all other resources retrieved from Google. This results in \RankSet{} containing a total of 2,617 queries and 46,881 resources.

To demonstrate the correctness of \MyModel's design and its applicability, we undertake an \textbf{ablation study}. \MyModel{} utilizes AdaRank as the underlying LTR algorithm with the expanded CS-DCG metric for optimization. To validate and examine how (i) the expansion of the optimization metric from the more traditional NDCG and (ii) the incorporation of \Bad{} as a sensitivity cost affect its overall performance, we compare \MyModel{} to AdaRank optimized with the standard NDCG metric. Each model is configured with variations that utilize each perspective as standalone features. To further contextualize the performance of \MyModel, we perform a \textbf{comparison} with a baseline and a state-of-the-art counterpart: (i) LambdaMART, a popular listwise LTR model that utilizes Multiple Additive Regression Trees \cite{friedman2001greedy}, with the overall ranking function being the linear combination of regression trees, and (ii) Korsce \cite{milton2020korsce}, a model designed to rank resources that align with 3$^{rd}$ to 5$^{th}$ grade educational curriculum, are comprehensible for children in that same grade range, are objective in content (i.e., not based in opinion), and are appropriate for the classroom.

To \textbf{measure} performance, we use NDCG@10 and Mean Reciprocal Rank (\textbf{MRR}). MRR seeks to spotlight the average ranking position of the first relevant item. In our case, we find it particularly important to position objectionable resources very low among retrieved results. Therefore, we also compute an alternative version of MRR, in which rather than accounting for the first relevant (ideal) item, we account for the position of the first objectionable item. We call this MRR$_{Bad}$, where a lower value indicates better performance. The significance of results is verified using a two-tailed student $t$-test with p$<$0.05; all results reported and discussed in the following section are significant unless stated otherwise.

\section{Results and Discussion}

We begin our evaluation of adapting LTR to children searching in the classroom by looking at how a known listwise LTR algorithm, AdaRank, optimized for a standard ranking metric (NDCG), performs when trained to rank according to our chosen perspectives. We train variations of AdaRank with each perspective, \Edu, \Read, and \Bad, each acting as a single feature. We refer to these variations with the suffixes -E, -R, and -O, resp. We train the same set of variations for \MyModel{} with the addition of ones that use the mixer to combine the \Edu{} and \Read{} perspectives into a single feature. We refer to these with the suffixes -M, where the mixed values are the only feature, and -MER, where the mixed values are used alongside the individual perspectives. Results of the experiments are presented in Tables \ref{tab:ranking-results-ablation} and \ref{tab:ranking-results-baselines}.

\begin{table}[h!]
\centering
\caption{Performance of \MyModel\ and baselines using \RankSet. * indicates significance w.r.t.\MyModel\ and bold indicates best performing for each metric.}
\label{tab:ranking-results-baselines}
{%
\begin{tabular}{@{}lllll@{}}
\toprule
\multicolumn{1}{c}{\textbf{Algorithm}} & \multicolumn{1}{c}{\textbf{\begin{tabular}[c]{@{}c@{}}Optimization \\ Metric\end{tabular}}} & \multicolumn{1}{c}{\textbf{NDCG}} & \multicolumn{1}{c}{\textbf{\begin{tabular}[c]{@{}c@{}}MRR \end{tabular}}} & \multicolumn{1}{c}{\textbf{\begin{tabular}[c]{@{}c@{}}MRR$_{Bad}$\end{tabular}}} \\ \midrule
LambdaMART                             & NDCG                                                                                          & \textbf{0.784}                    & \textbf{0.228}                                                                      & \textbf{0.081*}                                                                         \\
Korsce                                 & N/A                                                                                           & 0.753*                            & 0.209                                                                              & 0.163*                                                                                  \\
REdORank                               & nCS-DCG                                                                                       & 0.779                    & \textbf{0.228}                                                                      & 0.097                                                                                  \\ \bottomrule
\end{tabular}%
}
\end{table}

We first look at individual perspectives. As anticipated, AdaRank-O performed the worst, i.e., lower NDCG and MRR scores but higher MRR$_{Bad}$. We attribute this to AdaRank-O optimizing for the ``risk" perspective and thus learning to potentially prioritize the known bad resource above the known ideal. When optimizing on the ``reward" perspectives, AdaRank-E and AdaRank-R perform better than AdaRank-O. These models place objectionable resources around the 10$^{th}$ position according to MRR$_{Bad}$, while ranking the ideal ones around the 5$^{th}$ position, according to MRR (Rows 1--3 in Table {\ref{tab:ranking-results-ablation}}). This is indicative of these models learning to focus on the types of resources well-suited for our user group and environment. When considering all of the features together, AdaRank outperforms each of the individual variations, showcasing that the design choices for considering risk and reward perspectives in a re-ranking task are well-founded.

We surmise, however, that the AdaRank models are learning to rank objectionable resources lower as a beneficial side-effect of optimizing on the \Edu\ and \Read. To account for objectionable as an explicit signal of cost, and to balance that risk with the reward of the other perspectives, we turn to \MyModel, optimized for nCS-DCG.

For \MyModel-E, \MyModel-R, and \MyModel-O, we see similar performances to those of their AdaRank counterparts (Rows 5--7 and 2--4 in Table \ref{tab:ranking-results-ablation}, respectively). This further highlights that the perspectives matter. We posit that the interconnection of \Edu{} and \Read{} will serve as a beneficial composite signal for the relevance of resources. For this reason, we utilize the mixer to combine the two perspectives. Surprisingly, \MyModel-M performs worse in all metrics when compared to \MyModel-R and performs the same as \MyModel-E. To fully investigate whether this combined perspective could provide value to the re-ranking, we created \MyModel-MER. Lending credence to the idea of incorporating a combined perspective, \MyModel-MER outperformed each of the individual perspective variations. While this variation performed significantly better than \MyModel{} in terms of MRR$_{Bad}$, it performed worse for the other two metrics. This highlights that the explicit consideration of a sensitivity cost factor, alongside multiple perspectives of relevance, has beneficial effects on re-ranking resources for children searching in the classroom.

\begin{figure}[h]
    \centering
    \includegraphics[width=0.5\linewidth]{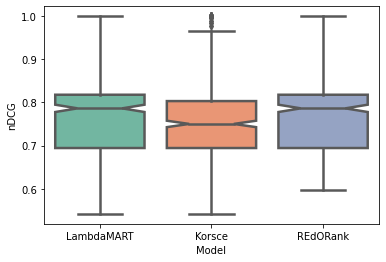}
    \caption{NDCG@10 for different re-ranking models using \RankSet.}
    \label{fig:ranking-box-plot}
\end{figure}

The results so far have shown that the design for \MyModel{} is well-founded. To attain a better understanding of how \MyModel{} performs, we also compare it to both a state-of-the-art counterpart, Korsce, and a baseline LTR algorithm, in LambdaMART. The results of these two models ranking the resources in \RankSet{} can be seen in Table \ref{tab:ranking-results-baselines}. We see that \MyModel{} performs significantly better than Korsce for all metrics. This is visually represented in Figure \ref{fig:ranking-box-plot}. We attribute the difference in performance to the fact that Korsce ranks in a pointwise, weighted objective manner. That is, for each resource, each perspective score is multiplied by an empirically determined weight and then added together to create the ranking score. In contrast, \MyModel{} learns a single dynamic weight that accounts for each perspective simultaneously as opposed to individually. LambdaMART learns to rank by optimizing on pairwise comparisons of documents. Surprisingly, LambdaMART performs significantly better than \MyModel{} for the \RankSet. While this was unexpected, as listwise LTR algorithms have been shown to be more effective when applied to Web search \cite{cao2007learning,tax2015cross}, we attribute the discrepancy in performance to the structure of the dataset. \RankSet{} only contains a single ideal resource, which a pairwise algorithm is more likely to ``locate" by nature of directly comparing documents. On the other hand, \MyModel{} is more likely to miss the ideal resource as it does not explicitly compare each resource to every other one but rather considers their relevance in a relative manner within the list. In real-world scenarios, where more than one ideal resource is likely to be in a single list, a listwise approach is better suited to the re-ranking task.

Going back to our RQ, given its visibly higher lower bound on NDCG@10 over its counterparts (Figure \ref{fig:ranking-box-plot}), its successful performance regarding ranking known educational and readable resources high in the rankings, and its expected generalizability to real-world re-ranking scenarios, we consider the design of \MyModel{} to be an appropriate model for providing re-ranking to search systems supporting children's online inquiry activities in the classroom.

\section{Conclusions, Limitations, and Future Work}
\label{ch:Conclusion}

\MyModel{}, the novel re-ranking strategy presented in this manuscript advances Information Retrieval for Children--centered on the design, development, and assessment of strategies that enable children's online information discovery. Given the broad ranges of children's search skills \cite{gossen2013specifics}, and inquiries children turn to SE for, we explicitly scoped our work to focus on children ages 6--11 using Web search tools in the classroom context.

Responding to \textbf{findings} reported in the literature that highlight children's propensity to focus on top-ranked results \cite{gossen2013specifics,gwizdka2017analysis}, as well as the manner in which SE handle children's queries \cite{azpiazu2017online,anuyah2019empirical}, e.g., offering resources children cannot comprehend, \MyModel{} examines resources retrieved by commercial SE and prioritizes them in a manner that those best suited for the context and user group at hand are ranked higher. In turn, \MyModel{} serves as a means to ease SERP exploration when children interact with the mainstream SE they are known to favor \cite{gwizdka2017analysis,azpiazu2017online}. To do so, \MyModel{} appraises resources based on three perspectives: \Edu, \Read, and \Bad. By combining all of these perspectives, \MyModel{} ensures that SE can better respond to children's search behavior by balancing the risk and reward value of resource content. An in-depth analysis of \MyModel{} revealed that a multi-perspective LTR model is an effective solution to re-rank resources for children in the classroom. The experiments conducted demonstrate that the deliberate inclusion of perspectives connected to a particular user group and environment can enhance model performance in re-ranking resources retrieved from a mainstream SE. 


We identified \textbf{limitations and pathways} for further research came to light. \ReadFormula{} was designed and evaluated on its applicability to English language resources. However, considering the vastness of the web and its diverse domains of information, conducting similar empirical investigations involving different domains like legal or medical, and exploring multilingual readability formulas, could offer valuable insights for various research fields. \MyModel{} leverages \Read{} as an internal feature, which only looks at text resources to estimate their readability. In the future, we plan to expand this perspective to consider other estimation methods that account for the presence of additional media elements, e.g., images and charts on web pages. Another limitation is the lack of consideration of a user's prior knowledge of a subject. Future work investigating the connection between pre-existing topical knowledge and readability estimation can bridge this gap and further align supporting tools such as \MyModel{} with their target user groups.
When exploring objectionable resources, we followed existing state-of-the-art approaches and treated all categories in \ObjCat{} as unquestionably objectionable. However, children do not necessarily require a one-size-fits-all solution. This is why we suggest increasing the granularity of \ObjCLF{} in identifying objectionable content based on specific age groups. Promising offline evaluations lead us to pursue further studies on the performance of \MyModel{} in a realistic environment. The next steps include a user study involving the examination of children's search behavior when using a search system with and without \MyModel.

Outcomes from this work have \textbf{implications} for researchers investigating children's Web search. \MyModel{} is a step towards adapting mainstream SE to classroom use, focusing on specific perspectives to inform relevance gain. It is worth researching the benefits of combining additional relevance signals beyond text, such as the origin or authorship of a resource. Such factors contribute to the credibility of a resource. Unfortunately, children are known not to judge the credibility of online resources \cite{hamalainen2020promoting}, making credibility a valuable extra perspective to bring into the fold for re-rankers. This can be achieved quickly and effectively using the mixer strategy employed by \MyModel{}, which enables the simultaneous aggregation of multiple scores into a single one. While this mixer is currently used for ``reward" perspectives, it can be replicated for ``risk" perspectives. For example, children have difficulty identifying misinformation \cite{pilgrim2021fake}, which may lead them to perceive misinformation as credible. By extending the mixer to include perspectives beyond \Bad, such as misinformation, \MyModel{} can prioritize resources that are based on accurate information \cite{landoni2023does}.

Ongoing research in Human-Computer Interaction has explored the impact of visual elements of a SERP on children's search behavior \cite{aliannejadi2021children,allen2021engage}. \MyModel{} provides further avenues of exploration regarding identifying resource types and visual elements that can serve as visual cues. For instance, adding a small book icon with a number to indicate the reading level of a resource or a schoolhouse icon to indicate educational value can be explored. Integrating visual elements that align with the ranking process can enhance the transparency of search systems, providing users with insights into how a particular system operates. This can impact the ease of use and understandability of a system. Additionally, such visual elements can benefit users learning to search.  Over time, the visible connection between ranked resources and search queries can help users improve their query formulation.

\section*{Acknowledgments}
Work partially funded by NSF Award \#1763649.

\bibliographystyle{unsrt}  
\bibliography{references}  

\end{document}